\documentclass[fleqn,12pt,twoside]{article}
\usepackage{espcrc1}

\usepackage{graphicx}
\usepackage[figuresright]{rotating}

\newcommand{\farcs}{\mbox{$.\!\!^{\prime\prime}$}}

\newcommand{\AmS}{{\protect\the\textfont2
  A\kern-.1667em\lower.5ex\hbox{M}\kern-.125emS}}

\title{\emph{CHANDRA} Observations of X-Ray Jet Structure on kpc to Mpc Scales}

\author{D. A. Schwartz\address{Harvard-Smithsonian Center for
Astrophysics \\ 60 Garden St., Cambridge, MA 02138, USA}
\thanks{This work was supported in part by NASA contract
NAS8-39073 to the \emph{Chandra} X-ray Center, NASA grant GO2-3151C to
SAO, and SAO SV1-61010 to MIT. E.S.P. acknowledges support from NASA
LTSA grant NAG5-9997.}, H. L. Marshall\address{Massachusetts Institute of
Technology \\ NE80-6031, Cambridge, MA 02139, USA}, B. P. Miller\addressmark,
D. M. Worrall\address{University of Bristol, Dept. of Physics \\
Tyndall Ave., Bristol BS8 1TL, United Kingdom},
M. Birkinshaw\addressmark, J.~E.~J.~Lovell\address{CSIRO Australia
Telescope National Facility \\ P.O. Box 76, Epping, NSW 2121, Australia}, D. L. Jauncey\addressmark,
E. S. Perlman\address{University of Maryland, Baltimore Campus \\
Physics Department, 1000 Hilltop Dr., Baltimore, MD 21250, USA},
D. W. Murphy\address{Jet Propulsion Laboratory \\ 4800 Oak Grove Dr.,
Pasadena, CA 91109, USA}, R. A. Preston\addressmark}

\begin{document}

\maketitle

\begin{abstract}

With its exquisite spatial resolution of better than 0.5 arcsecond,
the \emph{Chandra} observatory is uniquely capable of resolving and
studying the spatial structure of extragalactic X-ray jets on scales
of a few to a few hundred kilo-parsec.  Our analyses of four recent
Chandra images of quasar jets interpret the X-ray emission as inverse
Compton scattering of high energy electrons on the cosmic microwave
background. We infer that these jets are in bulk relativistic motion, 
carrying kinetic powers upwards of 10$^{46}$ ergs s$^{-1}$ to distances 
of hundreds of kpc, with very high efficiency.

\end{abstract}

\section{INTRODUCTION}
From our complete snapshot sample of quasar jets \cite{mars03,schw03},
we discuss here the four objects with the best 
statistical significance. These four powerful radio
quasars span a redshift range from 0.591 to 1.455, and illustrate a range of
morphological and physical effects.

We will compute source sizes and luminosities for a flat, accelerating
cosmology, with H$_0$=65 km s$^{-1}$ Mpc$^{-1}$, $\Omega_m = 0.3$ , and
$\Omega_{\Lambda}=0.7$, and use the formulas given by
\cite{Pen99}. For these jets, an H$_0$=50, q$_0 = 0.5$ cosmology 
changes lengths by  $\le$ 8\% and luminosities by  $\le$16\%.

\section{MORPHOLOGIES}

\begin{figure}[t]
\includegraphics[width=16cm]{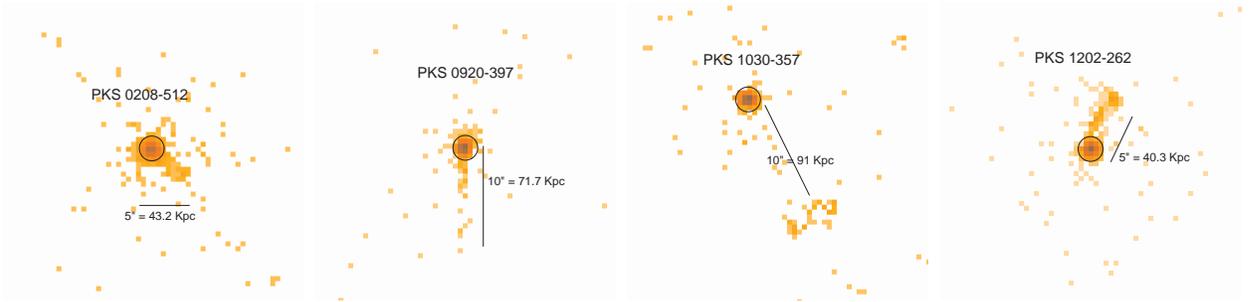}
\hspace{-.5in}
\caption{\label{xrayjets}X-ray counts in 0.5 to 7 keV range, binned in 0$\farcs$49 ACIS pixels.}
\end{figure}

The jets in these four objects display a variety of X-ray morphology
(Figure~\ref{xrayjets}).  The forms include the following: Straight
jets from core: PKS 0208-512, PKS 0920-397; curved jets and straight
jets which do not project to core: PKS 1202-262, (like PKS 0637-752
\cite{schw00}); gaps between core and jet: PKS 1030-357(?), (like
3C~273 \cite{mars01}, and PKS 0637-752); jet emerges from core,
disappears, resumes: PKS 0920-397, PKS 1030-357(?), (like Pic A
\cite{wils01}); X-ray to radio ratio roughly constant: PKS 1202-262,
(like PKS 0637-752); X-ray decreases away from core, radio increases:
PKS 0920-397, (like 3C~273); X-rays disappear when radio bends by
large angle: PKS~0208-512, PKS 1202-262, (like PKS~0637-752),
\emph{exception}: PKS 1030-357.  We suggest that all these features are
manifestations of small angular changes in jets which are beamed
toward our line of sight.

\begin{figure}[h]
\begin{minipage}{\textwidth}
\includegraphics*[width=1.5in]{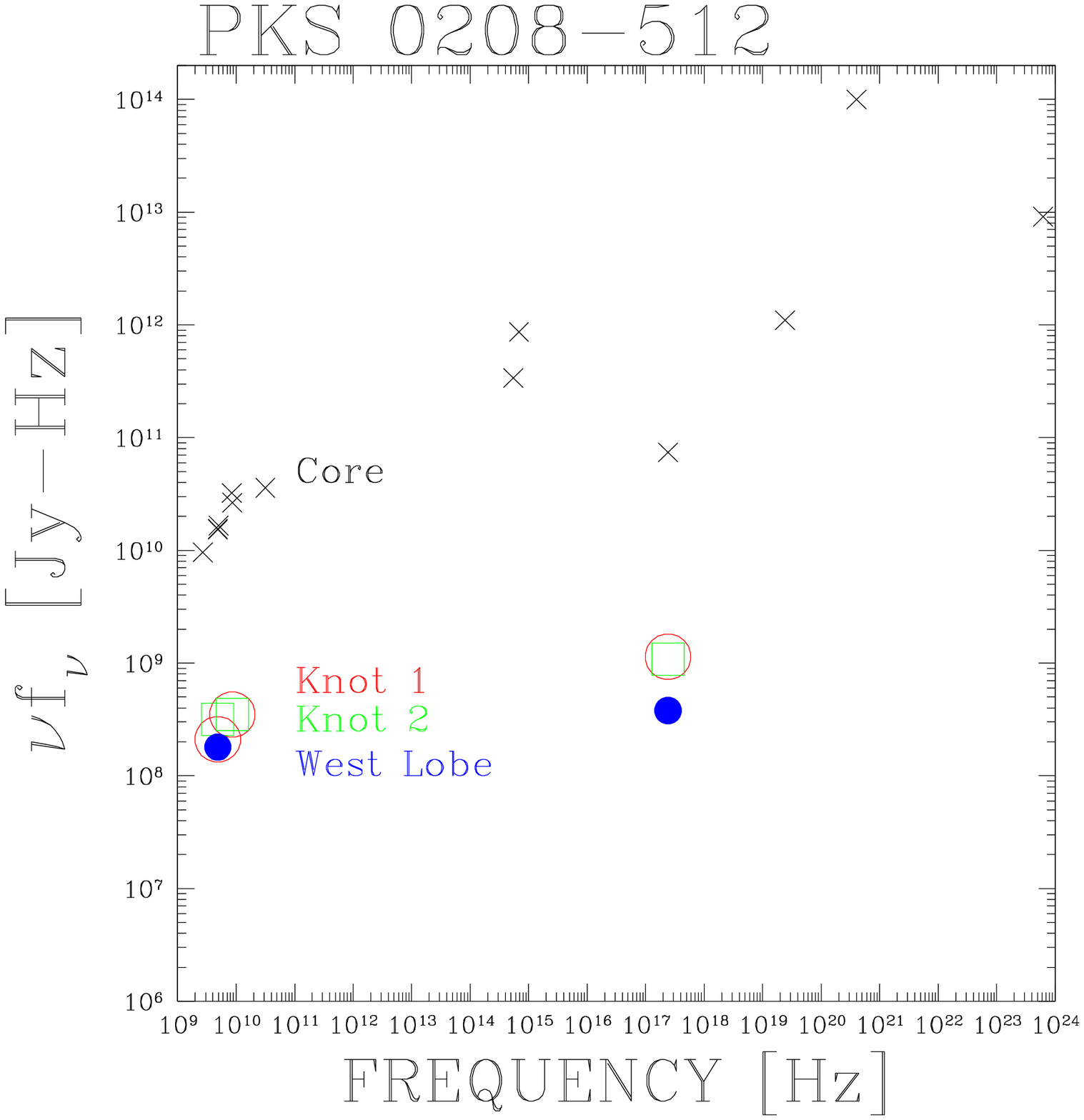}
\includegraphics*[width=1.5in]{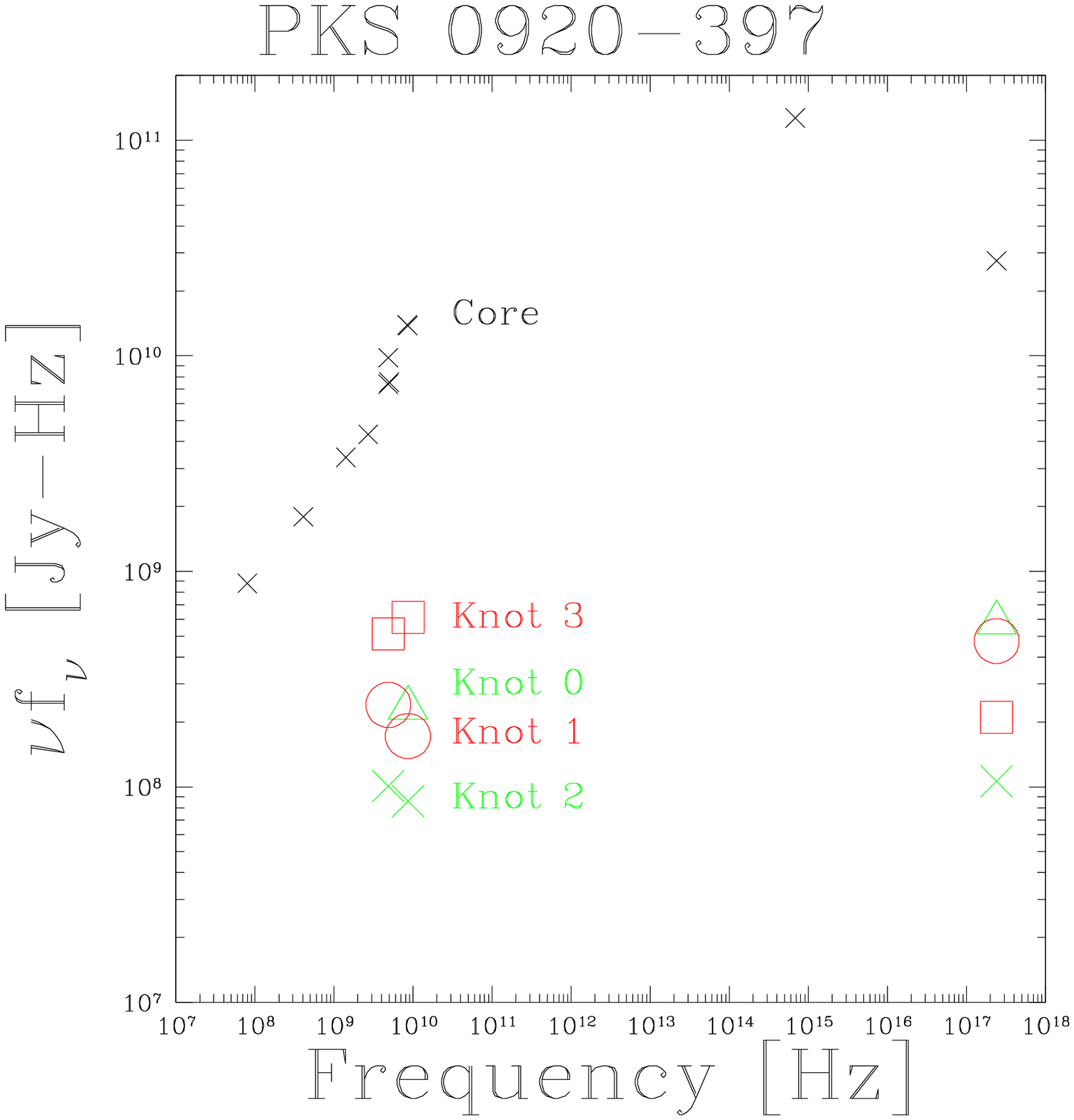}
\includegraphics*[width=1.5in]{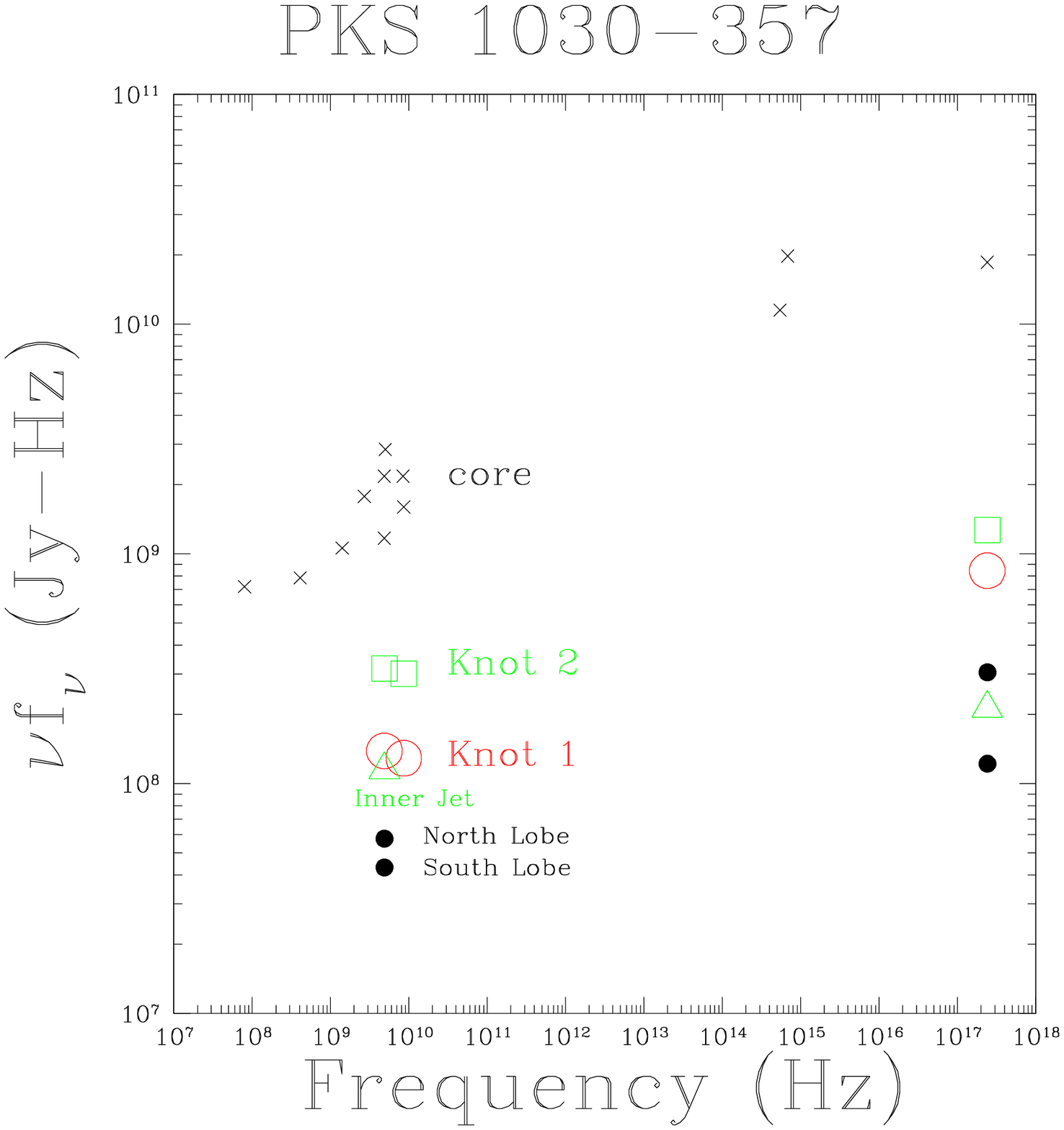}
\includegraphics*[width=1.5in]{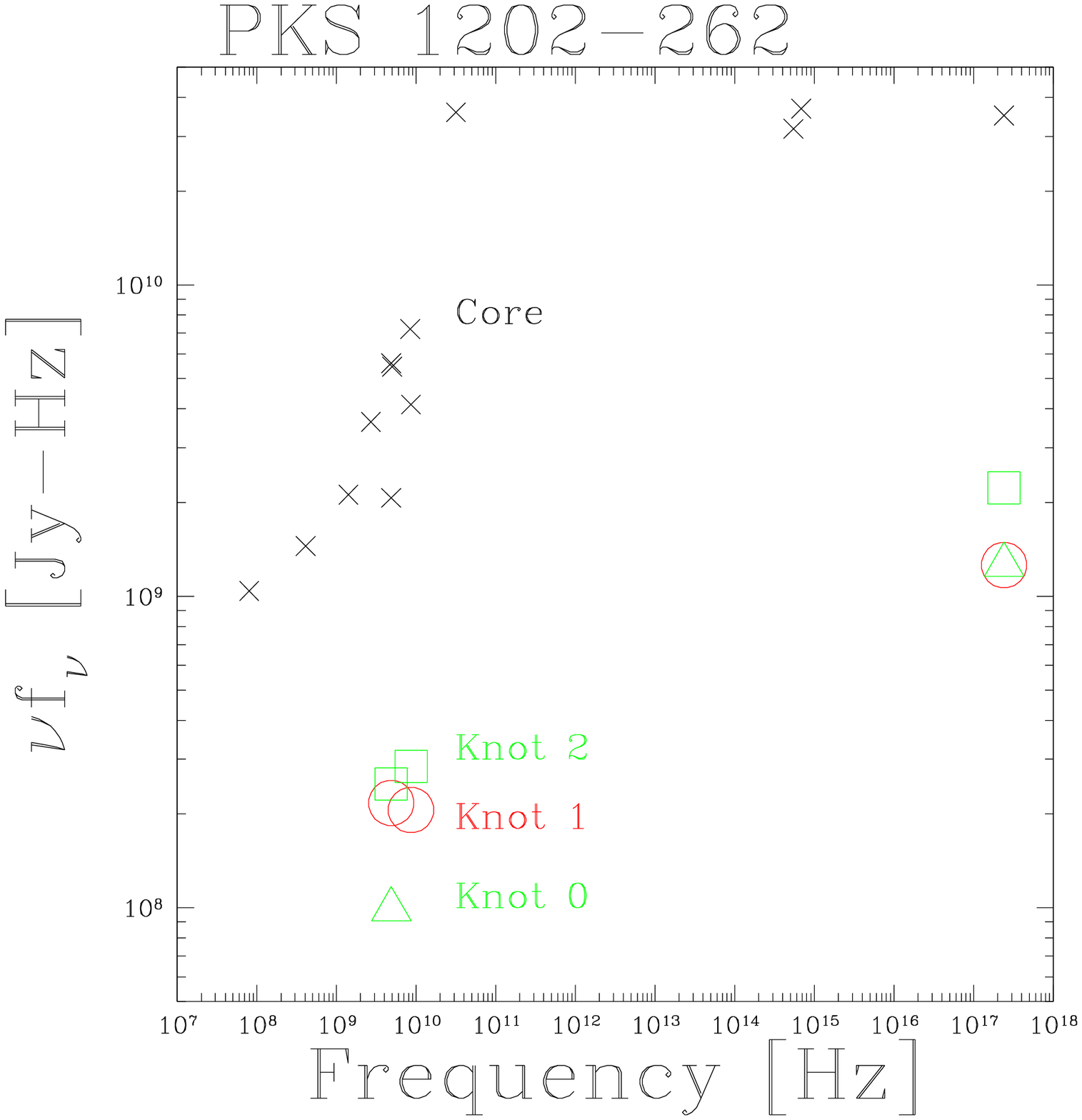}
\caption{\label{jetSED}SED for the quasar cores, and for each subregion of the four jets.  (Knot numbers increase away from the core.)}
\end{minipage}
\end{figure}

\section{EMISSION MECHANISMS}

We divide the jets into distinct regions, to compare the
changing X-ray and radio structure.  For each source we extract both
the X-ray and radio flux densities, and construct the broad-band
Spectral Energy Distributions (Figure~\ref{jetSED}).  (We
may have $\sim$20\% inaccuracies as we have not smoothed the
radio and X-ray data to correspond to the same angular
resolution.)  The jet radiative power is typically dominated by the
X-ray emission. Note the significant detection of X-rays from radio lobes
in PKS 0208-512 and PKS 1030-357.

For the gamma-ray blazar PKS 0208-512 we have measured \cite{mill02} an upper
limit to the optical emission of the jet which does not allow the X-rays
to be a simple extrapolation of the radio synchrotron emission.  For 
some  knots in the other sources, our two point radio spectral index
also argues against such extrapolation. We will assume that the X-ray emission from all the knots arises from
inverse Compton (IC) scattering by the same power law population of
electrons which emits the radio synchrotron radiation.

 The ratio of synchrotron to Compton flux densities at any frequency
is just the ratio of energy density of the magnetic field to the
energy density of the target photons \cite{felt66}.  In applying that
formula to the powerful X-ray jets, one typically cannot find a
credible source of target photons if one also assumes that the
magnetic field, B, and relativistic electrons are within an order of
magnitude of their equipartition values (e.g. \cite{schw00}).  This dilemma was
resolved by \cite{tave00,celo01} by exploiting the $\Gamma^2$
enhancement of the apparent CMB density in a frame moving with bulk
relativistic velocity with respect to the isotropic CMB frame
\cite{derm94}.

\begin{figure}[h]
\begin{minipage}{\textwidth}
\includegraphics*[width=1.5in]{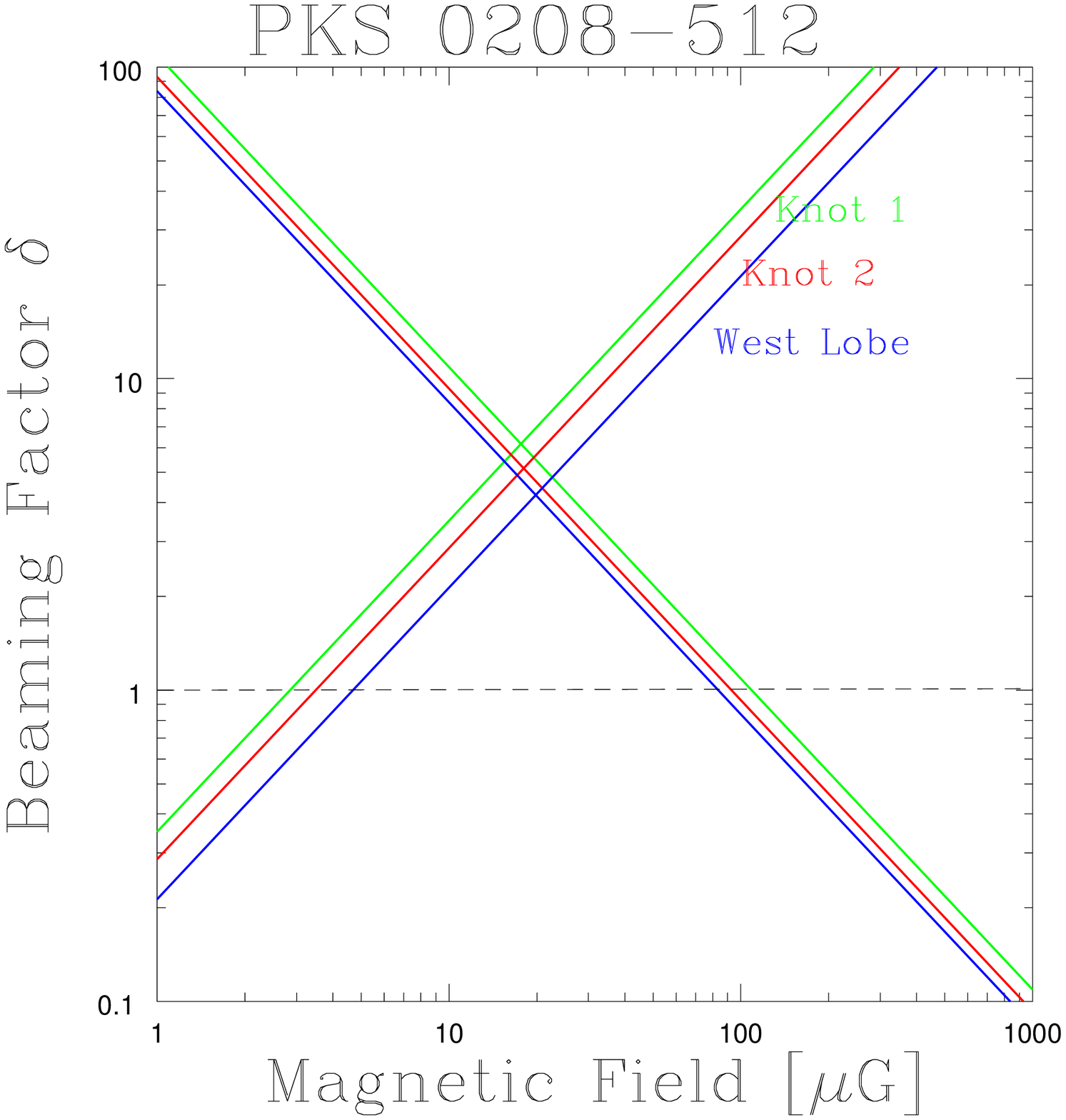}
\includegraphics*[width=1.5in]{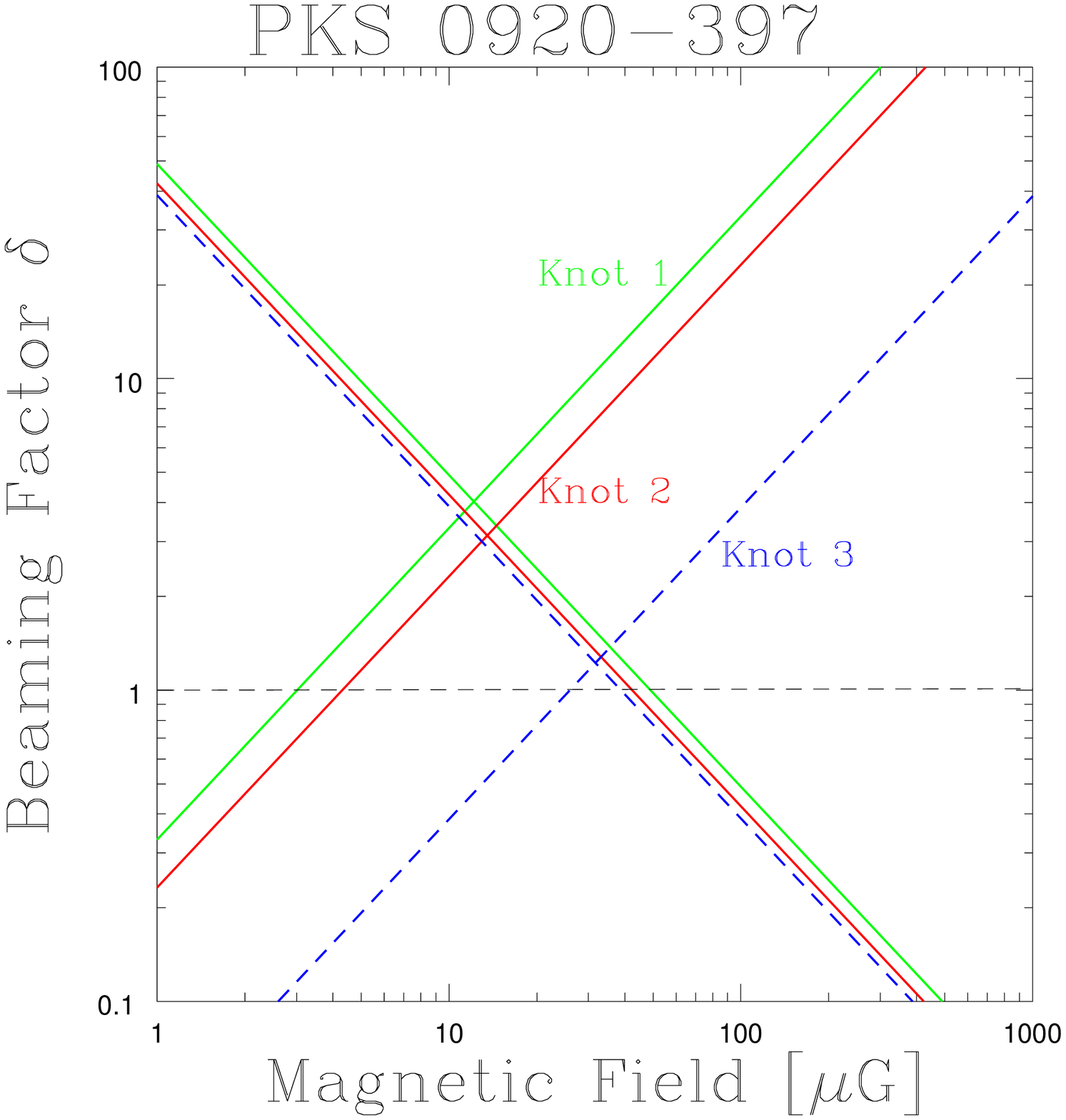}
\includegraphics*[width=1.5in]{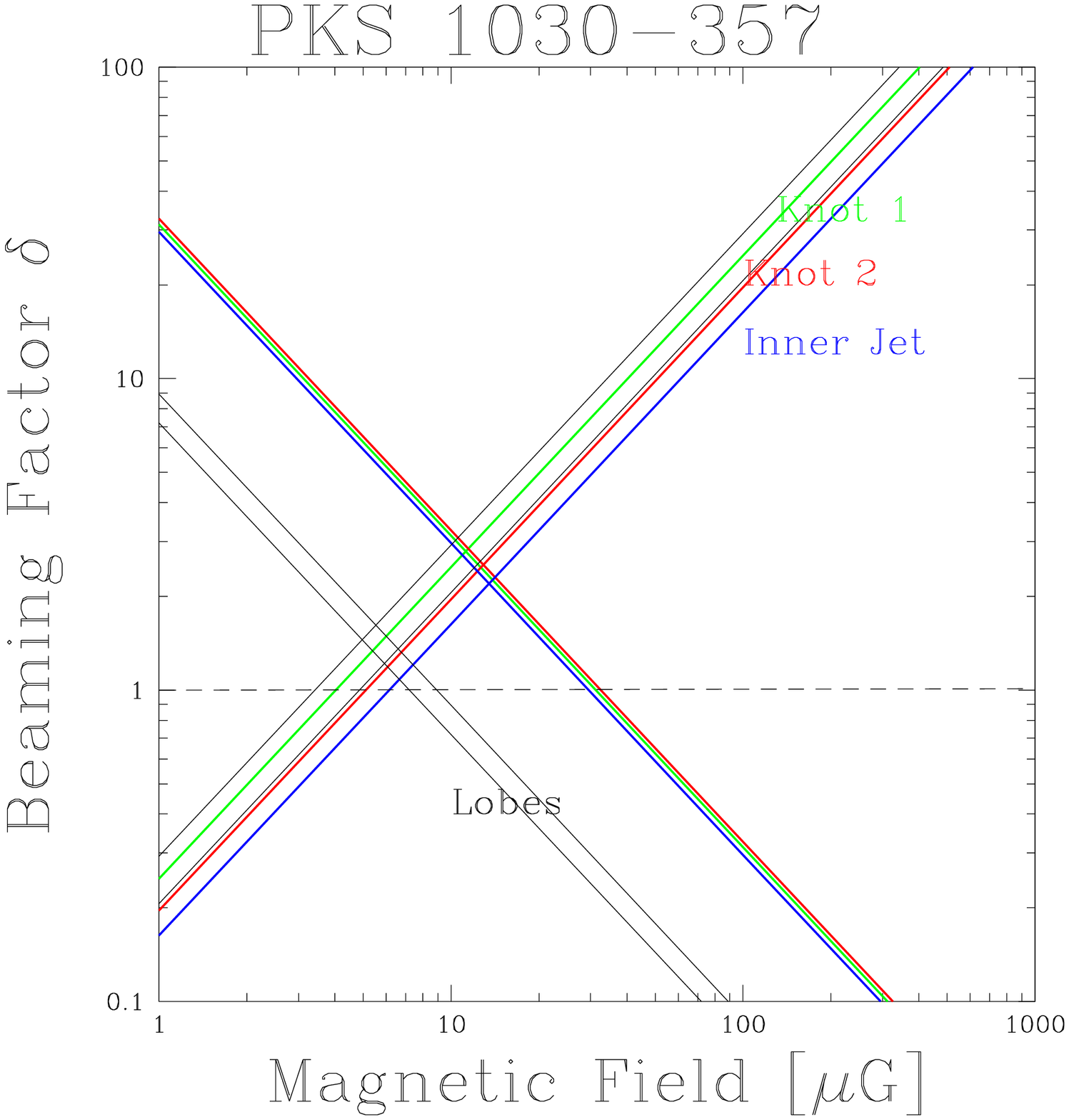}
\includegraphics*[width=1.5in]{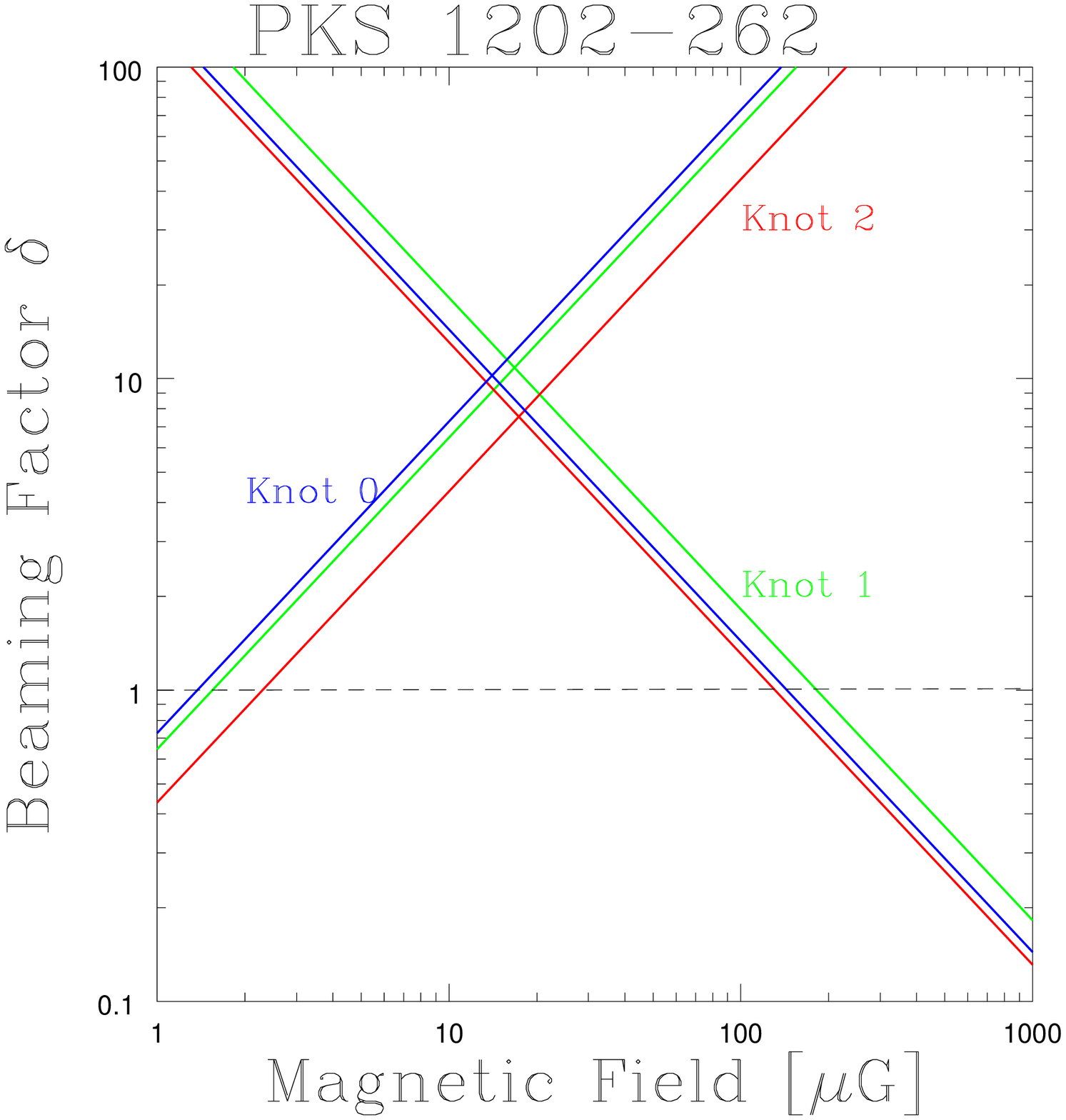}
\caption{\label{jettavec}The equipartition conditions
(lines decreasing with magnetic field) and the conditions for production of the
X-rays by IC/CMB (lines increasing with magnetic field), intersect to
give an estimate of the rest frame magnetic field and of the bulk relativistic beaming factor $\delta$.}
\end{minipage}
\end{figure}

A jet in bulk relativistic motion at an angle $\theta$ to our line of
sight is characterized by the beaming
factor $\delta=( \Gamma(1-\beta \cos \theta))^{-1} = ( \Gamma(1-\sqrt(1-1/\Gamma^2) \cos \theta))^{-1}$.
We have independent equations from the equipartition
condition, and from the condition that the electron population
producing the synchrotron radio emission also produces the X-rays via
inverse Compton scattering on the CMB.  In Figure~\ref{jettavec} the
intersections of these relations for each feature of each of the four
sources give  solutions for the unknowns $\delta$ and B.  We see a
pattern of magnetic fields of order 10 $\mu$G and relativistic beaming
factors 2 to 10 in the jets.  There is a hint that the $\delta$ values
decrease away from the cores; however, one needs to consider a factor
of a few uncertainty in B, since exact equipartition of energy need
not apply, and since the values we assume for the other parameters in
the equipartition calculation are not well determined.

\begin{table}[h] 
\caption{\label{table:1}\emph{Chandra} Observations of X-Ray Jets}
 
\begin{tabular}{cccccccccc}
\hline
PKS &
&
 &

Jet&
$<$B$>$&
 & 
 & 
 &
 &
Radiative \\

Name$^{a}$ &
z$^{a}$ &
L$_{\mathrm{x}}^{b}$ &

Fraction$^{c}$&
[$\mu$G]&
 $<\delta >$ & 
$\theta_{max}$ & 
L$^{d}$ &
P$_{\mathrm{j}}^{e}$ &
Efficiency$^{f}$ \\

\hline


0208-512  &0.999 &6.3 &0.035 &18
& 5.5 &10.5 &220  &12&0.002\\
 & & & & & & & \\
0920-397&0.591 &0.74 &0.062 & 13 &3.6& 16 &230  &1&0.005\\
 & & & & & & & \\
1030-357 &1.455 &3.5& 0.16 & 12 &  2.5 & 24 & 335  &4.5&0.012 \\
 & & & & & & & \\
1202-262&0.789&1.78 & 0.14  & 15  & 10.6  &5.4&470&5&0.005\\ 
\hline
\end{tabular}

\vspace{.07in}
$^{a}${NED, operated by JPL for NASA.}
$^{b}${Rest frame 2 --10 keV luminosity of quasar
core, in  10$^{45}$ergs s$^{-1}$.}
$^{c}${X-ray flux ratio, jet to core.}
$^{d}${Minimum length of X-ray jet, in kpc.}
$^{e}${Kinetic power of jet, 10$^{46}$ erg s$^{-1}$.}
$^{f}${Radiative power of jet, as a fraction of its kinetic power.}

\end{table}

We take a mean value of the magnetic field and of the beaming factors
to characterize the jet, as shown in Table~\ref{table:1}. For any
beaming factor $\delta$, the maximum angle between the jet and our
line of sight is $\theta_{max} =\cos^{-1} [(\delta -1/\delta)/\sqrt
(\delta^2 -1)]$.  From this maximum angle, and the measured projection
of the jet on the sky, we can compute the minimum intrinsic length of
each jet, as given in column 8. We estimate the kinetic power
transported by the jet as $\rm A \Gamma^2 c U$, where A is the cross
sectional area, $\Gamma$ is assumed equal to $\delta$, and U is the
total energy density in the jet rest frame (cf. \cite{ghis01}).  A
minimum power (column 9) results for an electron/positron jet, as
distinct from an electron/proton jet. These powers are typically
larger than the bolometric luminosity of the quasar,
(cf. Fig. ~\ref{jetSED}). The low efficiency, column 10, with which
these jets radiate their kinetic power is consistent with the ability
to transport energy from the black hole core to distant radio
lobes. Specifically in the case of PKS 0208-512, the power carried by
the jet on scales of 100 kpc is shown to be comparable to the power in
the pc scale jet as estimated by \cite{mara02}.

\end{document}